\documentclass[prl,a4paper,twocolumn,superscriptaddress,nobalancelastpage,nofootinbib]{revtex4-2}
\usepackage{xcolor}
\usepackage{amssymb,amsmath,bm}
\usepackage{graphicx}
\usepackage{tabularray}

\newcommand{\ket}[1]{\lvert #1\rangle}
\DeclareMathOperator{\rank}{rank}
\newcommand{\iprop}[2]{\mathbf{Int}(#1,#2)}

\begin{document}

\author{Norbert Schuch}
\affiliation{University of Vienna, Faculty of Mathematics, Oskar-Morgenstern-Platz 1, 1090 Wien, Austria}
\affiliation{University of Vienna, Faculty of Physics, Boltzmanngasse 5, 1090 Wien, Austria}

\author{Andr\'as Moln\'ar}
\affiliation{University of Vienna, Faculty of Mathematics, Oskar-Morgenstern-Platz 1, 1090 Wien, Austria}

\author{David P\'erez-Garc\'ia}
\affiliation{\mbox{Department of Applied Mathematics and Mathematical
Analysis, Universidad Complutense de Madrid, 28040 Madrid, Spain}}
\affiliation{Instituto de Ciencias Matematicas (CSIC-UAM-UC3M-UCM), 28049 Madrid, Spain}

\title{Simple Hamiltonians for Matrix Product State models}

\begin{abstract}
Matrix Product States (MPS) and Tensor Networks  provide a general
framework for the construction of solvable models. The best-known example
is the Affleck-Kennedy-Lieb-Tasaki (AKLT) model, which is the ground state
of a 2-body nearest-neighbor parent Hamiltonian.  We show that such simple
parent Hamiltonians for MPS models are, in fact, much more prevalent than
hitherto known: The existence of a single example with a simple Hamiltonian for
a given choice of dimensions already implies that any generic MPS with
those dimensions possesses an equally simple Hamiltonian. We illustrate
our finding by discussing a number of models with nearest-neighbor parent
Hamiltonians, which generalize the AKLT model on various levels.
\end{abstract}

\maketitle

\emph{Introduction.---}Solvable models play a key role in our understanding of quantum many-body
phenomena. This is particularly true in the study of strongly correlated
systems, where tools for a general analytical understanding are limited.
A paradigmatic example is the AKLT model~\cite{affleck:aklt-cmp}. It was
set forth as a model with a provable spectral gap, as predicted by Haldane for
integer spin chains~\cite{haldane:haldane-gap},  and formed one of the
earliest models realizing topological phenomena.  A key feature of the
AKLT construction is that it starts from a ground state wavefunction---the AKLT
state---from which one can construct an exact \emph{parent Hamiltonian},
and for which in turn properties such as a gap can be rigorously proven.
The AKLT construction was soon generalized by Fannes, Nachtergaele, and
Werner to a general class of 1D models~\cite{fannes:FCS}, now best known
as Matrix Product States (MPS). Later on, generalizations of these
constructions to more general settings and higher dimensions have been
devised, leading to the class of tensor network
models~\cite{cirac:tn-review-2021}, which nowadays form a general
framework for constructing and analyzing quantum many-body models.

A key feature of the AKLT state is that it possesses a particularly simple
parent Hamiltonian, given by a nearest neighbor interaction. 
Within the general framework of tensor network models and their parent
Hamiltonians, this property is very special: The theorems about parent
Hamiltonians with well-defined ground space yield Hamiltonians which generally act on a larger number of
sites, with the precise interaction range depending on specific parameters
used to set up the model (in particular, the dimensions of the involved
tensors).
For the AKLT model, those theorems do, in fact, only imply the existence
of a \emph{three}-site Hamiltonian, and manual postprocessing 
is required in order to show that it can be broken down into a two-site nearest
neighbor Hamiltonian.  However, what is the ``secret ingredient''
of the AKLT state which causes it to be the ground state already of a two-site
Hamiltonian? Given that we typically strive to devise models with
particularly simple (and thus realistic) Hamiltonians,
 it is highly desirable to trace down the origin of what
makes the AKLT so special. A similar question arises in two dimensions,
where yet again the AKLT model has a two-body nearest neighbor parent
Hamiltonian, while general parent Hamiltonian constructions yield
interactions acting on a large number of spins.

In this paper, we show that tensor network models with simple Hamiltonians
are much more common than anticipated. Concretely, what
we show is the following: Given a class of tensor networks---that is, an
underlying lattice together with a choice of dimensions for the physical
spins and the tensor degrees of freedom---the existence of a single
example with a simple parent Hamiltonian implies that this property is
generic, that is, any randomly chosen tensor network model in this class
will possess a well-behaved parent Hamiltonian with the same locality
structure.  Here, by well-behaved we mean that the Hamiltonian has a unique
ground state with a spectral gap above.  In particular, this has the
dazzling consequence that the fact that the AKLT model possesses a
very simple parent Hamiltonian implies that it is not special at all, by
the very fact of its existence.
More specifically, what we prove is that the set of tensor network models
without such a simple parent Hamiltonian is given by the space of zeros of a
real analytic function, and as such, it must either be the full space
(i.e., no tensor network model has this property), or it must be of
measure zero (i.e., the existence of a single example implies that 
the property is generic). Our proof applies to 
1D tensor network models with unique and degenerate
ground states as well as to a range of relevant classes of 2D models.
Motivated by the observation that simple parent Hamiltonians are abundant,
we search for new examples and present a range of new models with simple
parent Hamiltonians.

\emph{Matrix Product States.---}Let us for now focus on 1D systems.
For clarity, we will also focus on the translational invariant
(TI) setting, though our results directly generalize to the situation without
translational invariance. A (TI) MPS is constructed
from a $3$-index tensor $A^i_{\alpha\beta}$, $i=1,\dots,d$,
$\alpha,\beta=1,\dots,D$; alternatively, we can interpret $A^i$ as
elements of the space $\mathcal M_D$ of $D\times D$ matrices. The
MPS on $\ell$ sites with boundary condition $X$ is defined as
\begin{equation}
\label{eq:mps-def}
\ket{\Psi_\ell[X]} := 
\sum_{i_1,\dots, i_\ell=1}^d \mathrm{tr}[A^{i_1}\cdots
A^{i_\ell}\,X]
\ket{i_1,\dots,i_\ell}\ .
\end{equation}
With open boundary conditions (OBC), it spans the MPS space 
\begin{equation}
\mathcal S_\ell :=
\Big\{\,\ket{\Psi_\ell[X]}  \,\Big\vert X\in \mathcal M_D  \Big\}\ .
\end{equation}
For periodic boundary conditions (PBC), $\ket{\Psi[X=\openone]}$ defines
a TI MPS.
Clearly, $\mathrm{dim}\,\mathcal S_\ell\le D^2$. We say that an MPS is
\emph{normal} if there exists an $\ell$ such that $\mathrm{dim}\,\mathcal
S_\ell=D^2$, and call the smallest such $\ell$ its \emph{injectivity
length} $L_0$. Generic MPS are normal, and their injectivity length $L_0$ takes
the minimal possible value (the smallest $L_0$ such that $d^{L_0}\ge
D^2$)~\cite{perez-garcia:mps-reps,cirac:tn-review-2021,jia:wielandt-generic}.
In the following, we focus on normal MPS.

\emph{Parent Hamiltonians.---}Whenever $\mathcal S_\ell\subsetneq (\mathbb
C^d)^{\otimes \ell}$, we can define an $\ell$-site Hamiltonian
$h^\ell:=\openone-\Pi_{\mathcal S_\ell}$ (with $\Pi_{\mathcal S_\ell}$ the
orthogonal projector onto $\mathcal S_\ell$); when acting on part of a
spin chain, $h^{\ell}_i$ denotes $h^\ell$ acting on sites
$i,\dots,i+\ell-1$. Fix some $N\ge \ell$. $h^\ell$ is positive
semidefinite, and $h^\ell_i\ket{\Psi[X]}=0$; thus, $\mathcal S_N$ is
contained in the ground space of the OBC Hamiltonian $H^\ell_N =
\sum_{i=0}^{N-\ell} h^\ell_i$, and $\ket{\Psi[\openone]}$ in the ground
space of the corresponding PBC Hamiltonian.\footnote{Importantly, this
means that any ground state of $H_N^\ell$ is already a ground state of
each term $h^\ell_i$; that is, $H_N^\ell$ is \emph{frustration free}.}  We
thus have obtained a Hamiltonian together with a succinct description of
some of its ground states.  For this to give a meaningful solvable model,
we however should be able to characterize the \emph{entire} ground space
of $H^\ell_N$. The ideal scenario---which is the one we are after---is
when the ground space of $H^\ell$ is just given by the MPS itself, i.e.,
spanned by all $\ket{\Psi[X]}$, and thus equal to $\mathcal S_N$.  Since the
ground space of $H^\ell_N$ is the intersection of the ground spaces of
$h^\ell_i$, this is the case exactly if the \emph{intersection property}
$\mathbf{Int}(\ell,N)$, defined as
\begin{equation}
\label{eq:intersection-property}
\mathbf{Int}(\ell,N):\ 
\mathcal S_N \stackrel{!}{=} \bigcap_{i=0}^{N-\ell} 
    (\mathbb C^d)^{\otimes i} \otimes \mathcal S_\ell \otimes (\mathbb
C^d)^{\otimes (N-\ell-i)}
    =:\mathcal I^{\ell}_N
\end{equation}
holds for the given $N$.  Note that $\mathcal S_N\subset \mathcal
I^{\ell}_N$ by definition of $\mathcal S_\ell$; this implies that
$\mathbf{Int}(\ell,N)$ holds precisely if $\dim \mathcal I^\ell_N=D^2$.
An established fact about normal MPS is that for any $L\ge L_0+1$ and
$N\ge L+1$, $\iprop{L}{N}$ holds; that is, the ground space of the parent
Hamiltonian $h^L$, constructed on at least $L_0+1$ sites, is exactly given
by the MPS space $\mathcal S_N$~\cite{perez-garcia:mps-reps,cirac:tn-review-2021}. 
Moreover, once $\iprop{\ell}{L}$ holds for some $L\ge L_0+1$,
the ground state of the corresponding $\ell$-site PBC Hamiltonian on $N\ge
\mathrm{max}(L_0+\ell-1,L)$ sites is unique (and thus given by
$\ket{\Psi[\openone]}$)~\cite{perez-garcia:mps-reps,schuch:peps-sym,cirac:tn-review-2021}.\footnote{Any 
state in the ground space is both of the form $\ket{\Psi_N[X]}$ and its
translation by $L_0$ sites with boundary condition $Y$. Inverting the
joint $L_0$-length block yields $XA^{i_1}\cdots A^{i_{\ell-1}} = A^{i_1}\cdots
A^{i_{\ell-1}}Y$, which implies an exponentially degenerate OBC ground
space---cf.\ later---unless $X=Y=\openone$.}
Due to this implication, we will adopt the terminology ``$H_N^\ell$ has a unique
ground state'' to also describe the scenario where the OBC ground space equals
$\mathcal S_N$ for sufficiently large $N$. (This also implies that the
OBC ground state is unique in the bulk, since in normal MPS, 
the dependence on the boundary conditions decay exponentially
into the bulk~\cite{cirac:tn-review-2021}.)
Note that by \eqref{eq:intersection-property}, it follows that if $\iprop{\ell}{L}$
and $\iprop{L}{N}$ hold, also $\iprop{\ell}{N}$ holds;
thus, once we know that $\iprop{\ell}{L}$ for \emph{some} $L\ge
L_0+1$, it follows that the parent Hamiltonian with terms $h^\ell$ has a
unique ground state. Finally, once we have established uniqueness of the
ground state, is is known that the parent Hamiltonian always exhibits a
spectral gap above~\cite{fannes:FCS}.

Let us now discuss the AKLT model. It is given by a normal MPS with
$d=3$, $D=2$, with $A^i$ the Pauli matrices~\cite{cirac:tn-review-2021}.
Injectivity is reached for $L_0=2$, which is the generic behavior for 
this $d$ and $D$.
Since $\iprop{L_0+1}{N}$ always holds,  the AKLT MPS
precisely spans the ground space of the parent Hamiltonian constructed on
$L_0+1=3$ sites. On the other hand, $\dim {\mathcal S}_{L_0}=D^2=4$, but
$\mathcal S_{L_0}\subset (\mathbb C^3)^{L_0}\cong \mathbb C^9$; we can thus 
construct a non-trivial \emph{two-site} Hamiltonian $h^2 = \openone -\Pi_{S_2}$ for the AKLT state.
However, does this Hamiltonian still have a unique ground state? This will
be the case if it satisfies $\iprop{2}{L}$ for some $L\ge L_0+1=3$.  For 
the specific case of the AKLT model, it can be easily checked that 
this is indeed the case for $L=3$.  This establishes that the AKLT
state is the unique ground state of the two-site parent Hamiltonian $h^2$,
which is nothing but the well-known AKLT Hamiltonian.

\emph{Main theorem.---}The existence of a
non-trivial Hamiltonian acting on $L_0$ sites is based purely on dimensional arguments,
and works for any pair of dimensions $d$ and $D$ for which
$D^2=d^{\ell}$ does not have an integer solution $\ell$: Since
$D^2<d^{L_0}$, one can construct a parent Hamiltonian already on $L_0$
sites, rather than $L_0+1$. However, which conditions does the underlying
MPS have to fulfill such that the resulting Hamiltonian has the
intersection property $\iprop{L_0}{L}$ for some $L>L_0$, and thus a unique
ground state?

To analyze this, we consider when $\iprop{\ell}{L}$, cf.\
Eq.~\eqref{eq:intersection-property}, fails.
This is the case if and only if the ground space $\mathcal I^\ell_L$ of
$H^{\ell}_L$ is strictly larger than the ground space $\mathcal S_L$ of
$h^L$.  Since $h^L$ is a projector, this happens precisely when
\begin{equation}
\label{eq:fA-det-zero}
 f(A):= \det\left[H_L^\ell + (1-h^L)\right] \stackrel{!}{=}0\ ,
\end{equation}
which is a function of the MPS tensor $A$.

We thus need to analyze the zeros of $f$. As the domain of $f$, 
we consider the space $\mathcal G$ of MPS tensors $A$ for which $\dim
\mathcal S_{\ell}=D^2$, i.e., which are injective on $\ell$ sites.
If we define $P_A:X\mapsto \ket{\Psi_{\ell}[X]}$ [with $A$ the MPS tensor,
cf.\ Eq.~\eqref{eq:mps-def}] and $T_A=P_A^\dagger P_A^{\phantom\dagger}$,
we have that $\mathcal G\equiv\{A:\rank
T_A=D^2\}$. Since $P_A$ and thus $T_A$ are continuous and the rank is
lower semi-continuous, it follows that $\mathcal G$ is an open set.  Let
us now show that $\mathcal G$ is moreover (path-)connected, i.e., it is a
\emph{domain}, based on an argument by Szehr and
Wolf~\cite{szehr:cpmap-connected-components}.  To this end, consider
$B,C\in\mathcal G$.  Define $A(z)=(1-z)B+zC$, $z\in\mathbb C$, and let $\tilde
T(z):=P_{A(0)}^\dagger P^{\phantom\dagger}_{A(z)}$.  Since $\rank
P_{A(0)}=D^2$, we have that $\det \tilde T(0)\ne 0$, and moreover,
$\det\tilde T(z)$ is a polynomial in $z$, and thus can only have a finite
number of zeros. This implies that we can find a path $z\in \mathbb C$
which interpolates from $z=0$ to $z=1$ such that along the entire path,
$\det\tilde T(z)\ne 0$, which implies $\rank P(z)=D^2$ and thus yields
that the path which interpolates between $A,B\in\mathcal G$ is contained
in $\mathcal G$---that is, $\mathcal G$ is path-connected.

Let us now consider the structure of $f(A)$.  
$T_A$ has full rank on $\mathcal G$, and since $\mathrm{Im}\,P_A =
\mathcal S_\ell$,
we have
\begin{equation}
h_\ell=\openone - \Pi_{{\mathcal S}_\ell}= 
\openone - P_A^{\phantom\dagger} T_A^{-1} P_A^\dagger \ .
\end{equation}
Since $P_A$ and $T_A$  are polynomials in $A$ and $\bar A$, we find that
$h_\ell$,  and thus $H^\ell_L$, is a real analytic function. Moreover,
since injectivity on $\ell$ sites implies injectivity also on $L>\ell$
sites, $h^L$ is real analytic as well.  Thus, $f(A)$ is a real analytic
function on $\mathcal G$.  The MPS where the intersection property fails
are thus the zeros of the real analytic function $f(A)$ on the domain
$\mathcal G$ [Eq.~\eqref{eq:fA-det-zero}], and they must therefore either be the full space, or a set
of measure
zero~\cite{mityagin:zeros-real-analytic-function}.\hspace{\fill}$\blacksquare$

\begin{figure}
\includegraphics[width=\columnwidth]{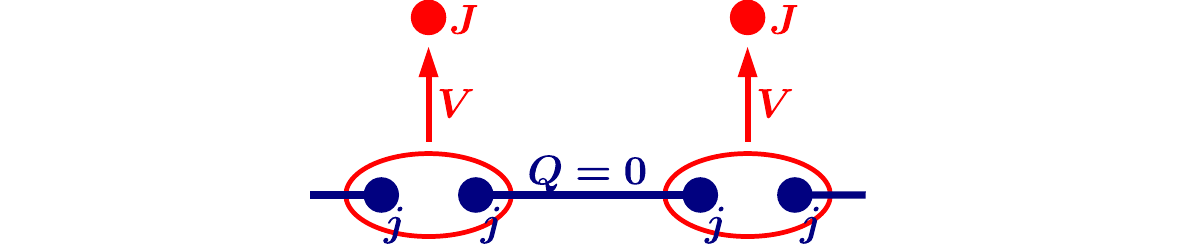}
\caption{\label{fig:aklt}
AKLT-type models are constructed by projecting pairs of virtual spin-$j$
particles onto their physical spin $J$ subspace by an isometry $V$.
Virtual spins between adjacent sites are placed in a singlet, $Q=0$
(``bond'').  Overall, the construction shown can be seen as a map
$P:X\to\ket{\Psi_2[X]}$ from the two virtual spin-$j$ particles 
at the boundary to the two physical spin-$J$ particles, cf.\
Eq.~\eqref{eq:mps-def}.  We also consider generalizations where
the bond has $Q\ne0$.}
\end{figure}

\emph{Examples.---}We now look at examples of MPS models which
have a parent Hamiltonian which acts on the injectivity length $L_0$ and 
possesses the intersection property, and thus a unique ground state.
We focus on $L_0=2$, since this gives models with two-body
nearest neighbor Hamiltonians; in that case, the interesting range of
dimensions is $D<d<D^2$  ($D<d$ implies that generically, $L_0\le2$ and 
$h^2$ is non-trivial, and $d<D^2$ rules out $L_0=1$).  Following our 
main theorem, each of the subsequent examples, by its mere existence,
implies that for generic MPS models with the corresponding choice of
dimensions, the $2$-site Hamiltonian has a unique ground state.

The first example is the AKLT model. It has $d=3$, $D=2$, with $A^i$
the Pauli matrices. Equivalently, it can be constructed in terms of
$\mathrm{SO(3)}$ spins, by starting from a chain of
virtual spin $j=\tfrac12$ singlets which are then pairwise projected onto their
joint $J=1$ space to yield the physical spin $J$,
 as shown in Fig.~\ref{fig:aklt}. The AKLT model has $L_0=2$,
and has a non-trivial $2$-body parent Hamiltonian $h^2$. $h^2$  can be
constructed by observing that the physical spin $J$ on two 
consecutive sites
can take values $1\otimes 1 =0\oplus 1 \oplus 2$, while the state prior to
the projection only takes values $\tfrac12\otimes0\otimes\tfrac12=0\oplus
1$, as the central spin-$\tfrac12$s form a singlet ($L=0$).
The ground space of $h^2$ is thus precisely spanned by the subspaces with spin 
$0\oplus 1$, i.e., $h^2$ equals the projector onto the joint spin-$2$
subspace. It can be straightforwardly checked that the model indeed possesses
the intersection property $\iprop{2}{3}$, and thus, the two-body 
Hamiltonian has a unique ground state.

A family of generalized AKLT models, set forth in
Ref.~\cite{fannes:FCS}, is obtained by starting from a chain of spin-$j$
singlets and projecting pairs of those onto their spin-$J$ space, see
Fig.~\ref{fig:aklt}. For $J=2j$, the resulting model has $L_0=2$, and as
proven in Ref.~\cite{fannes:FCS}, $\iprop{2}{3}$ holds; thus, the two-body
parent Hamiltonian again has a unique ground state.  This yields a family
of examples with $d=2J+1=4j+1=2D-1$ for all $D\ge 2$. Furthermore, any
example for some $d$ also gives examples for all $d'>d$, by
trivially embedding the physical system into the larger space. Thus, we
find that for all $d\ge 2D-1$, generic MPS are the unique ground states of
two-body parent Hamiltonians.

\begin{table}
{
\footnotesize
\setlength{\tabcolsep}{1.5ex}
\begin{tabular}{c|c||cccccc}
$D$ & $d$ & $L=2$ & $L=3$ & $L=4$ & $L=5$ & $L=6$  & $L=7$ \\
\hline
\hline
   $3$   &    $4$   &    $9$  &    $9$  &  $\cdots$ \\
\hline
   $4$   &    $5$   &   $16$  &   $35$  &   $31$  &   $16$  & $\cdots$ \\
   $4$   &    $6$   &   $16$  &   $16$  & $\cdots$ \\
\hline
   $5$   &    $6$   &   $25$  &   $84$  &  $229$  &  $450$  &  $181$  &   $25$  \\
   $5$   &    $7$   &   $25$  &   $25$  &   $\cdots$ \\
\hline
 $6$ & $7$    &  $36$   &  $161$  &  $659$  &  $2520$ &  $9073$  & $30751$  \\
 $6$ & $8$    &  $36$   &  $64$   &   $36$  & $\cdots$ \\
 $6$ & $9$    &  $36$   &    $36$  & $\cdots$ 
\end{tabular}
}
\caption{\label{table:int-dim}
Ground space degeneracy $\dim \mathcal I_L^2$ of the $2$-site parent
Hamiltonian of generic MPS with virtual and physical dimension $D$ and $d$ on
blocks of $L=2,\dots,7$ sites.}
\end{table}

\emph{Generic Matrix Product States.---}What about MPS models with $D+1\le d < 2D-1$---are
they also generically unique ground states of their two-site parent
Hamiltonians? To assess this
question, we numerically test random MPS. The $\dim \mathcal
I_L^2$ obtained are listed in Table~\ref{table:int-dim} (recall that $\dim\mathcal
I_L^\ell\ge D^2$, with equality exactly when the intersection property is
reached). Let us go through the results row by row.  First, we find that
also for $D=3$, $d=4$, $\iprop{2}{3}$ holds. But for $D=4$ and $d=D+1=5$, we
observe something surprising. The dimension $\dim \mathcal I^2_L$ of the
ground space at first increases with
$L$, yet later decreases to reach $D^2=16$ at $L=5$---that is,
$\iprop{2}{5}$ holds and thus and the $2$-body Hamiltonian has a unique
ground state, but in order to detect this, we need to consider a $5$-site block. The
same behavior is seen for $D=5$ and $d=D+1$, but now we even have to consider a
$7$-site block. As we increase $d$ for $D=4$ or $D=5$, the intersection
property is again reached immediately for $L=3$. 
In fact, since an example for some $d$ also provides one for $d'>d$, we
find that whenever generically $\dim\mathcal I^2_L=D^2$ for some $d$ and
$L$, the same must also hold generically for $d'>d$ for the same $L$.  
Finally, for $D=6$ and $D=d+1$, the rapid growth of the dimension of
$\mathcal I_L^2$ makes it impossible to determine numerically whether
$\dim \mathcal I_L^2$ eventually decreases again and possibly reaches
$D^2$.

That $\dim \mathcal I_L^2$ initially increases with $L$ might come as a
surprise. However, it is in fact unavoidable, as can be seen
from parameter counting. $h^2$, embedded in an $L$-site chain, imposes
$(d^2-D^2)d^{L-2}$ constraints (orthogonality to all vectors not in the ground
space).  Thus, 
\begin{equation}
\label{eq:dim-lowerbnd}
\dim\mathcal I^\ell_L \ge d^L - (L-1)(d^2-D^2)d^{L-2}\ ,
\end{equation}
with equality if the constraints from different Hamiltonian terms $h^2_i$
are independent.  Depending on $D$ and $d$, this can indeed
give non-trivial lower bounds on $\dim \mathcal I^\ell_L$. In fact, we
numerically observe that for $L=3$, this bound is tight for generic instances (as
those in Table~\ref{table:int-dim})---that is, the constraints in $h^2$ on
two consecutive sites are independent (which is plausible given the
absence of any reflection symmetry), as long as the r.h.s.\ in
\eqref{eq:dim-lowerbnd} is larger than $D^2$; we have tested this
for all $D+1\le d < 2D-1$ up to $D=30$.
On the other hand, the bound \eqref{eq:dim-lowerbnd} is no longer tight
for $L\ge 4$ (that is, the constraints are no longer independent, which
must happen at some $L$ since the bound eventually goes below $D^2$).

\emph{Further examples.---}What are further concrete examples beyond the
aforementioned spin-$J$ AKLT models, and in particular, can we find some
with a $D/d$ ratio closer to $1$?  One possibility is to generalize the
AKLT construction, Fig.~\ref{fig:aklt}, to $j\le J<2j$. However, for tensors
with injectivity length $L_0=2$, the resulting $2$-site Hamiltonian no
longer has a unique ground state, as observed in Ref.~\cite{fannes:FCS}:
The reason is that injectivity implies that the ground space on two sites,
$J\otimes J=0\oplus\cdots\oplus 2J$, precisely carries \emph{all} possible spins
obtained from the virtual spins at the boundary, $j\otimes 0\otimes j =
0\oplus\cdots\oplus 2j$.
This condition is, however, also met by any generalized AKLT state with
smaller virtual dimension $j'<j$, and thus, the ground space also contains
those states~\cite{fannes:FCS}.  In fact, we can choose a different spin
$j_b$ on every bond $b$, as long as $j_b\otimes j_{b+1}$ contains the
physical $J$-spin, resulting in an exponentially degenerate ground
space.\footnote{If we choose $j_{b_e}=j$ on every even bond $b_e$,
injectivity across this bond implies that states with a different choice
$\{j_{b_o}\}_{b_o}$ on the odd bonds $b_o$ are all linearly independent.}

However, there is still the possibility that some generalized AKLT
models---we term those \emph{exceptional AKLT models}---do not have
injectivity length $L_0=2$, a case not considered in
Ref.~\cite{fannes:FCS}. Indeed, all we know is that on $2$ sites, the
construction in Fig.~\ref{fig:aklt} gives a map of the form
$P = \sum_{S=0}^{2j} w_S P_S$, where $P_S$ is the unique isometry mapping the
spin-$S$ space of $j\otimes j$ at the boundary to the spin-$S$ space in $J\otimes J$, and
the $w_S$ are \emph{some} weight obtained from summing the Clebsch-Gordan
coefficients, 
which for some particular choices of $j$, $J$, and $S$ can happen to be zero. 
This is e.g.\ the case for $J=2$, $j=3/2$, where $w_2=0$.
 In this case, we find that the injectivity length is
$L_0=4$, while $\dim \mathcal S_{2}=11$, $\dim \mathcal S_3=15$, and the ground
space of the $2$-site parent
Hamiltonian $H^2_N$ on $N\ge2$ sites is precisely given by $\mathcal S_N$---in
particular, the ground space of the $2$-site PBC Hamiltonian is unique for
system sizes $N\ge 5$. Therefore, this constitutes an
example with a well-behaved parent Hamiltonian acting on \emph{less}
than $L_0$ sites.  Concretely, the Hamiltonian $h^2$ of
this new solvable gapped spin-$2$ model must be positive precisely on the
subspaces with total spin $S=2$ and $S=2j+1=2J=4$, and zero otherwise; this gives
rise to a one-parameter family of nearest-neighbor Hamiltonians $\lambda
H_a + (1-\lambda) H_b$ (where positivity requires $\lambda<60/53$) with
representatives \newcommand{\sds}{\vec J_i\cdot \vec J_{i+1}}
\begin{align*}
H_a &= \sum_{i=1}^N -\sds+\tfrac{91}{900}(\sds)^3 + \tfrac{11}{900}(\sds)^4
\\
H_b &= \sum_{i=1}^N (\sds)^2 + \tfrac{11}{30} (\sds)^3 + \tfrac{1}{30} (\sds)^4
\end{align*}
which all have the same unique ground state and a spectral gap
~\cite{tu:SOn-mps-gs,tu:soN-vs-su2-mps}.
Further examples of this kind are obtained for $J=5$,
$j=3$~\cite{tu:exact-spinchain-rg-su2}, and
for $J=9$, $j=5$, which again have two-body Hamiltonians with a unique
ground state, and satisfy $\iprop{2}{4}$.

Let us now turn back to examples with $L_0=2$. We have seen
that such $\mathrm{SO}(3)$-invariant models beyond 
generalized AKLT models with $J=2j$
are not possible; hence, we instead turn towards models
with a $\mathrm{U}(1)$ symmetry.  Specifically, we modify the AKLT model
by
replacing the bond singlet by a state with spin $Q\ne 0$ and $Q_z=0$,
see Fig.~\ref{fig:aklt}.
Within this class, we find a large number of choices for $j$ and $J$ with
$L_0=2$, and with a $\mathrm{U}(1)$-symmetric nearest neighbor Hamiltonian with a
unique ground state, which we list in the Appendix; for instance,
two cases with a $j/J$ ratio closer to $1$ are $j=7/2$, $J=5$, $Q=4$,
which has $\iprop{2}{5}$, and $j=4$, $J=6$, $Q=4$, which has
$\iprop{2}{4}$.

\emph{Generalizations.---}Our main theorem---that small parent Hamiltonians,
if even a single example exists, are generic---immediately generalizes to a range of
other settings. First, the theorem is not restricted to MPS with
minimal injectivity length, but applies to the space of all
MPS which are injective at some fixed length $L$. Second, we can consider
MPS with block-diagonal but otherwise normal tensors $A^i$ (i.e.,
block-injective MPS), as long as we fix the block structure: All we have
to do is to modify the proof such as to show connectedness of the space of
block-injective MPS tensors with the given block structure, by using that
$\tilde T(z)$ has full rank on the corresponding space.
Finally, the same proof applies to injective tensor networks in higher
dimensions, or on general graphs, and it can yet again be generalized to
tensor networks which are block-injective on a given fixed subspace.
Here, such a block structure can relate to states with long-range order
such as a GHZ state~\cite{cirac:tn-review-2021}, to topologically ordered states
(G-injective~\cite{schuch:peps-sym},
MPO-injective~\cite{sahinoglu:mpo-injectivity}, or
Hopf-injective~\cite{molnar:weak-hopf-mpo} tensor networks), or to
semi-injective PEPS where injectivity at corners is restricted to a
subspace with some symmetry~\cite{molnar:quasi-injective-peps}, such as
the 2D AKLT state. Some examples are the Majumdar-Ghosh chain,
the kagome Resonating Valence Bond state, or Kitaev's quantum double
models~\cite{majumdar:majumdar-ghosh-model,kitaev:toriccode,verstraete:comp-power-of-peps,schuch:rvb-kagome,schuch:peps-sym}.

\emph{Beyond unique ground states.---}We have already seen that there are
cases of MPS with $L_0=2$ where the $2$-site Hamiltonian has an
exponentially degenerate ground space: The AKLT-class states with $2j>J$.
More generally, examples with exponentially degenerate ground space can be
obtained from tensors which satisfy $XA^i=A^iY$ for some $X\ne Y$: By
placing, or not placing, $X$ on every second link we obtain linearly
independent states (due to injectivity) which are all ground states of the
$2$-body Hamiltonian (as $X$ can be replaced by a $Y$ on the adjacent
link). Since such $(X,Y)$ form an algebra, one chooses
$X^2=X$, $Y^2=Y$ to only have one such pair. For any such choice,
$XA^i=A^iY$ is a linear equation, and we find that its solutions
generically have injectivity length $L_0=2$.\footnote{This can be naturally generalized to
larger interaction lengths by demanding 
$XA^{i_1}\cdots A^{i_{\ell}}= A^{i_1}\cdots A^{i_{\ell}}B$, as well as to
correlated actions $A^iXA^j = \sum_k Y_kA^iA^jZ_k$.}
An entirely different ground space structure is given by MPS
where $\mathcal S_2$ contains the span of some other MPS $B^i$, in which
case also that MPS will be a ground state; in fact, this completely
characterizes all cases where the parent Hamiltonian has a bounded ground
state degeneracy~\cite{matsui:gapped-ham-vs-mps}.

\begin{acknowledgments}
\emph{Acknowledgments.---}We acknowledge helpful discussions with JM
Landsberg (who in particular pointed out that parameter counting can give
non-trivial lower bounds to the ground space degeneracy) and Andreas
Klingler.  This research was funded in part by the Austrian Science Fund
FWF (Grant No.\ \href{https://doi.org/10.55776/COE1}{10.55776/COE1},
\href{https://doi.org/10.55776/P36305}{10.55776/P36305}, and
\href{https://doi.org/10.55776/F71}{10.55776/F71}), the European Union --
NextGenerationEU, and the European Union’s Horizon 2020 research and
innovation programme through Grant No.\ 863476 (ERC-CoG SEQUAM).
This work was in part supported by the Spanish Ministry of Science and
Innovation MCIN/AEI/10.13039/501100011033 (grants CEX2023-001347-S,
PID2020-113523GB-I00, PID2023-146758NB-I00), and by Comunidad de Madrid
(grant TEC-2024/COM-84-QUITEMAD-CM).  For open access purposes, the
authors have applied a CC BY public copyright license to any author
accepted manuscript version arising from this submission.
\end{acknowledgments}

\onecolumngrid

\vspace*{-0.5cm}
\subsection*{Appendix A: $\mathrm{U}(1)$-symmetric generalizations of the
AKLT model}

In this appendix, we summarize
$\mathrm{U}(1)$-symmetric generalizations of the AKLT construction where
the singlet bond is replaced by a bond state with spin $Q$ and magnetic
quantum number $Q_z=0$, and where $j$ and $J$ denote the virtual and
physical spin, respectively, see also Fig.~\ref{fig:aklt}.
Table~\ref{table:u1} lists all values of $Q$ for $j=1,\dots,4$ and
$J=2,\dots,8$ which have $L_0=2$ and for which we can verify that the
intersection property $\iprop{2}{\ell}$ is reached for some $\ell$. 
The corresponding minimal value of $\ell$ is also given in the table. 

We observe that in general the value of $\ell$ is independent of $Q$, and only depends on $j$ and $J$.
 The one exception is $j=4$, $J=4$, where $Q=2$ shows a
deviating behavior, as given in the table. Note that $j\ge J$ is ruled out
since in that case we cannot have $L_0=2$ (or only with a trivial parent
Hamiltonian for $j=J$), while for $J>2j$, the AKLT construction is
ill-defined as $J$ is not contained in $j\otimes j$; the corresponding
cells are greyed out. For $j$, $J$ combinations with empty white cells,
there is no $Q$ such that $L_0=2$ and $\iprop{L_0}{\ell}$ can be verified
to hold for the attainable $\ell$.  Finally, note also that $Q\le 2j$ must hold, and that $Q=0$
corresponds to the $\mathrm{SU}(2)$-invariant generalized AKLT
construction.

There are also two combinations $j$ and $J$ in the table which for $Q=0$
correspond to the exceptional AKLT models discussed in the main text,
that is, which are not injective for $L_0=2$, yet whose $2$-site parent
Hamiltonian has a unique ground state. Further analysis shows that this
exceptional behavior also occurs for $j=\tfrac32$, $J=2$, $Q=2$, as
well as $j=3$, $J=5$, $Q=3$ and $j=5$, $J=9$, $Q=4$.

\begin{table}[h]
\begin{tblr}{hline{1,2,9}={2pt,solid},hline{3-8}={1pt,solid},
	vline{1,2,9}={2pt,solid},vline{3-8}={1pt,solid},
	rows={abovesep=4pt},
	cells={valign=m},
	cell{1}{2-8}={halign=c},
	cell{3}{2}={halign=c},
	cell{6}{5}={halign=c},
	cell{2}{3-8}={gray},
	cell{3}{4-8}={gray},
	cell{4}{5-8}={gray},
	cell{5}{6-8}={gray},
	cell{6}{7-8}={gray},
	cell{7}{8}={gray},
	cell{4}{2}={gray},
	cell{5}{2}={gray},
	cell{6}{2-3}={gray},
	cell{7}{2-3}={gray},
	cell{8}{2-4}={gray},
	cell{2-8}{1}={valign=m} 
}
  & $J=2$ & $J=3$ & $J=4$ & $J=5$ & $J=6$ & $J=7$ & $J=8$
\\
$j=1$  &  {$Q=0,2$\\$\ell=3$}
\\
$j=\tfrac32$ &   {exceptional\\model} &  {$Q=0,1,2,3$\\$\ell=3$}
\\
$j=2$ & & & {$Q=0,2,4$\\$\ell=3$}
\\
$j=\tfrac52$ & & & {$Q=2,3,4,5$\\$\ell=4$} & {$Q=0,\ldots,5$\\$\ell=3$}
\\
$j=3$ & & & {$Q=4,6$\\$\ell=6$} & {exceptional\\model} & {$Q=0,4,6$\\$\ell=3$}
\\
$j=\tfrac72$ & & & & 
{$Q=1,\ldots,7$\\$\ell=5$} & 
{$Q=2,\ldots,7$\\$\ell=4$} & 
{$Q=0,\ldots,7$\\$\ell=3$}
\\
$j=4$ & & & & & 
{$Q=2,4,6,8$\\$\ell=4$} & 
{$Q=2(\ell=4),4,6,8$\\$\ell=3$} & 
{$Q=0,2,4,6,8$\\$\ell=3$}
\end{tblr}
\caption{\label{table:u1}
Generalized AKLT model with $\mathrm{U}(1)$ symmetry, where the bond is
placed in a spin-$Q$ state with $Q_z=0$, cf.\ Fig.~\ref{fig:aklt}. The
table lists all values for $Q$ for the given $j$, $J$ for which the MPS is
injective on two sites ($L_0=2$) and has a unique ground state; $\ell$
gives the minimal length for which the intersection property
$\iprop{2}{\ell}$ holds. Greyed out cells are forbidden, as is $Q>2j$.
}
\end{table}

\subsection{Appendix B: Numerics for ground space degeneracies and MPS
structure of ground space}

In this appendix, we explain how the ground space degeneracy of
frustration free Hamiltonians can be computed efficiently even for large
system sizes, such as in Table~\ref{table:int-dim}. Specifically, we will
see that the computational resources scale polynomially with the ground
space degeneracy rather than the size of the physical Hilbert space. As a 
noteworthy side result, we will see that the ground space of frustration free
Hamiltonians can always be parametrized by part of a half-infinite
(non-translational invariant) MPS, where the bond dimension at each
position equals the ground space dimension to the left (assuming we
grow the system towards the right). All these results hold for any
frustration free Hamiltonian on a line, regardless of how it has been
constructed and what properties its ground space has. 

For simplicity, we restrict to $2$-body Hamiltonians $H=\sum_{i=1}^{N-1}
h_i$, where $h_i$ acts on sites $i$ and $i+1$, and $N$ can be infinity.
We denote the ground space of $H_L = \sum_{i=1}^{L-1}h_i$ ($L\le N$) by
$\mathcal K_L$.

To understand that the ground space is described by an MPS which can be
computed efficiently, with the bond dimension equal the ground state
degeneracy, we can proceed inductively. Assume the ground space $\mathcal
K_L$ of $H_L$ is spanned by an MPS
\begin{equation}
\raisebox{-1em}{\includegraphics[scale=0.95]{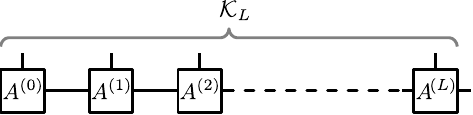}}
\label{eq:app:KL-MPS}
\end{equation}
where the tensors $A^{(k)}$, interpreted as maps from the right to the top
and left index, are isometries.
This implies that the entire MPS in Eq.~\eqref{eq:app:KL-MPS} (as well as any of its
left parts until site $s\le L$) describes an isometry from the open bond
on the right to the physical space, and the ground space dimension is
precisely the bond dimension $D_L$.  

The induction hypothesis is certainly true for $L=2$: We can obtain the
MPS by computing the ground space (i.e., kernel) of the $2$-site
Hamiltonian, construct an isometry $\hat V = \sum_{ij} V_{ij}^\alpha
\ket{i,j}\langle\alpha\rvert$ which spans said ground space, and write
this as an MPS
\begin{equation}
\label{eq:app:A-from-V}
\raisebox{-2.5em}{\includegraphics[scale=0.95]{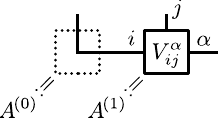}}
\end{equation}
spanning the ground space,
where the tensor $A^{(0)}$ implements the identity between virtual and
physical degree of freedom.

In order to obtain the ground space on $L+1$ sites starting from
Eq.~\eqref{eq:app:KL-MPS}, we need to take 
$\mathcal K_L\otimes \mathbb C^d$ and intersect it with the ground space
of the next Hamiltonian term $h_L$, that is, we need to find an isometry
$V$ which spans the solution space of the equation
\begin{equation*}
\raisebox{-3.3em}{\includegraphics[scale=0.95]{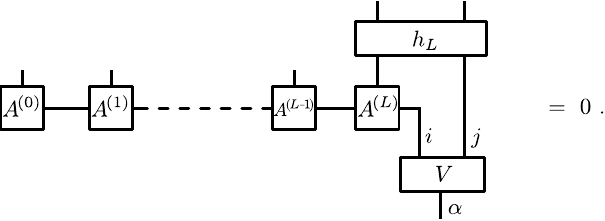}}
\end{equation*}
Using that the MPS on the left (from $A^{(0)}$ until $A^{(L-1)}$) is an
isometry, this is equivalent to finding the $V$ which spans the solution
space of 
\begin{equation*}
\raisebox{-3.3em}{\includegraphics[scale=0.95]{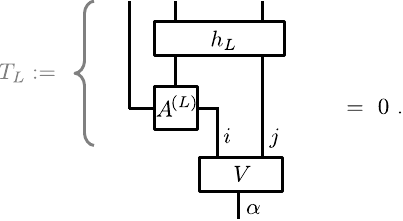}}
\end{equation*}

We now see that the dimension of the operator $T_L$ whose null space we need to
find is governed by the bond dimensions $D_L$ and $D_{L-1}$, rather
than the exponentially growing physical Hilbert space dimension $d^L$, and
an isometry $V$ spanning this solution space can be computed with
resources scaling polynomially in $D_L$, $D_{L-1}$. $V$ then readily
gives the next MPS tensor $A^{(L+1)}$, analogously to Eq.~\eqref{eq:app:A-from-V}.

Let us note that for growing bond dimensions (i.e., ground space
degeneracies), a larger system size can be
potentially solved if one grows the ground space both from the left and
the right, and in a final step connects the two MPS in the middle by projecting onto the
central Hamiltonian term.

\end{document}